\documentclass{article}
\usepackage{graphics}
\usepackage{graphicx}
\usepackage{xcolor}
\usepackage{amssymb}
\usepackage{float}
\usepackage{hyperref}
\usepackage{amsmath}
\newtheorem{thm}{Theorem}[section]

\begin{document}

\title{A rigorous multi-population multi-lane hybrid traffic model and its mean-field limit for dissipation of waves via autonomous vehicles}
\author{Nicolas Kardous\thanks{These authors contributed equally to this work}, Amaury Hayat\thanks{These authors contributed equally to this work}, Sean T. McQuade, Xiaoqian Gong,\\ Sydney Truong, Tinhinane Mezair, Paige Arnold, Ryan Delorenzo, \\Alexandre Bayen, Benedetto Piccoli\thanks{authors are from: Department of Electrical Engineering and Computer Sciences, University of California Berkeley; Department of Mathematics, Rutgers University Camden; CERMICS, Ecole des Ponts Paristech; Center for Computational and Integrative Biology, Rutgers University Camden; School of Mathematical and Statistical Sciences, Arizona State University}}

\maketitle

\begin{abstract}
In this paper, a multi-lane multi-population microscopic model, which presents stop and go waves, is proposed to simulate traffic on a ring-road.
Vehicles are divided between human-driven
and autonomous vehicles (AV). 
Control strategies are designed with 
the ultimate goal of using a small number of AVs (less than 5\% penetration rate) to represent Lagrangian control actuators that can smooth the multilane traffic flow 
and dissipate the stop-and-go waves.
This in turn may reduce fuel consumption and emissions.\\

The lane-changing mechanism is based on three components that we treat as parameters in the model: safety, incentive and cool-down time.
The choice of these parameters in the lane-change mechanism is critical to modeling traffic accurately, because different parameter values can lead to drastically different traffic behaviors.
In particular, the number of lane-changes and the speed variance are highly affected by the choice of parameters.\\

Despite this modeling issue,
when using sufficiently simple and robust controllers for AVs, the stabilization of uniform flow steady-state is effective for any realistic value of the parameters, and ultimately bypasses the observed modeling issue.\\

Our approach is based on accurate and rigorous mathematical models, which allows a limit procedure that is termed, in gas dynamic terminology, mean-field. In simple words, from increasing the human-driven population to infinity, a system of
coupled ordinary and partial differential equations are obtained. Moreover, control problems also pass to the limit, allowing the design to be tackled at different scales.
\hfill\break%
\end{abstract}

\section{Introduction}
\label{intro}
Traffic flow displays various instabilities
at high densities, and this is known as a congested phase.
Such instabilities may grow into persistent
stop-and-go waves and travel upstream to the flow of traffic. This phenomenon is especially observed on highways, and was reproduced in experiments \cite{jiang2014traffic,Sugiyamaetal2008}. There has been a large effort in the recent decades to explain the origin of the such waves, and its potential link to the nonlinearity of the system (see for instance \cite{cui2017stabilizing,orosz2009exciting}).
Waves may be generated by network features
(bottlenecks, ramps etc.) as well
as by drivers' behavior (lane changing, strong
breaking, etc.). These waves are responsible for traffic inefficiencies
and increased fuel consumption.
There is a wide literature on traffic modeling by different research communities, we refer the readers to \cite{MR3553143,Helbing01,Kerner_1999} for general discussions.

Traditional traffic management techniques
include variable speed advisory and
variable speed limits.
However, the technological advancements in terms of measurements and autonomy allowed the use of a Lagrangian approach
using autonomous vehicles as Lagrangian actuators that are sparse along the road network.
A number of studies addressed the problem
of dampening waves to smooth traffic using
autonomous vehicles, 
both in simulation
\cite{PhysRevE.69.066110,GUERIAU2016266,orosz2010traffic,Hedrick94,TALEBPOUR2016143,7362183}
as well as in experiments
\cite{WU201982,STERN2018205}.
The achieved results also showed that at low
penetration (around 5\%), traffic can
be smoothed to a great extent
in terms of fuel economy (reduction up
to 40\%). The effects of connected automated vehicles on traffic patterns were investigated in \cite{avedisov2020impacts,oh2020safe}. In particular, long-range feedback may benefit traffic flow and that car-following models with delay are able to replicate the experimental results. 
The results from the experiments were mostly
in a confined setting and only used one lane, with
controls designed from first principles
and control-theoretic methods,
\cite{7995897,delle2019feedback}.

The present paper aims at designing 
rigorous mathematical models
and control algorithms for a multi-lane setting in a ring road. 
More precisely, we design a multi-population model with human-driven and autonomous vehicles. The microscopic dynamics is described by a Bando-Follow-the-Leader model, proven to present instabilities (generating in particular stop-and-go waves), and tuned to experimental data.
The lane-changing mechanism is mainly
based on MOBIL (“Minimizing Overall Braking Induced by Lane Changes”) \cite{treiber2009modeling}
and includes: safety, incentive and cool-down time. Safety poses constraints on acceleration/deceleration of vehicles,
incentive is based on the potential for higher acceleration in a new lane and cool-down time allows lane-changing only after a certain amount of time from the last lane change.
The resulting dynamics is of a hybrid nature and heavily depends on the choice of parameters for these three mechanisms.

In particular,
we show in this article that the quantitative but also qualitative behavior of the dynamics (persistent instabilities or not ; number of lane-change; speed variance) highly depends on the parameters of the lane changing mechanism. 
Despite the variability of traffic patterns,  we show that we can still design
simple control algorithms, which are robust and can stabilize traffic with a low penetration rate for any choice of parameters
(in a physically relevant parameters' space). \textcolor{black}{Finally, we show that a collaborative driving approach, where a minority of vehicles would have "good human behavior", would also bring some stability to the system.}

The hybrid model and control strategies are based on accurate mathematical analysis.
This type of rigorous mathematical hybrid system, in turn, has been shown to allow a limiting procedure,
called mean-field, with the population of human-driven cars sent to infinity.
The limiting controlled dynamics couples
a partial differential equation for the 
human-driven car density with
a controlled hybrid system
for the autonomous vehicles. Optimal control problems have also been shown to be compatible with the limiting procedure. This microscopic model and microscopic control is a first-step to allow control strategies to be designed
for the limiting dynamics and at different scales.

The paper is organized as follows: in Section \ref{sec:1} we discuss existing models in the literature and present our models. In Section \ref{sec:2} we study the influence of the lane-changing parameters on the traffic behavior. In Section \ref{sec:3} we see how autonomous vehicles can be used to smooth traffic instabilities, in particular stop-and-go waves in a robust way (with respect to the change in lane-changing parameters). In Section \ref{sec:4} we discuss the potential of collaborative driving where both collaborative drivers and non-collaborative drivers co-exists. In Section \ref{sec:futureworks} we discuss the potential of this model to allow a mean-field procedure.

\subsection{Existing traffic models in the literature}
\label{sec:1}
A multilane model is typically composed of two components : longitudinal dynamics for each lane and a lane-change mechanism. Due to the different scales that can be represented in vehicular traffic, one can also classify traffic models by modelling longitudinal dynamics for each lane into two typical categories: micro-models and macro-models. For general discussions about traffic models at different scales, we refer to the survey papers \cite{piccoli2009vehicular,bellomo2011modeling,albi2019vehicular}. 

{\itshape a) Micro-model}\\
There are many different micro scale traffic models. We show several continuous time models here that are well-understood and used regularly: the Intelligent Driver Model (IDM), the Bando model, and the Follow The Leader (FTL) model. For the IDM, introduced in \cite{treiber2000congested}, the longitudinal dynamics for a lane are written:

\[
\begin{cases}
\dot{x}_i = v_i,\\
\dot{v}_i = a\bigg(1-\left(\frac{v_i}{v_0}\right)^\delta - \left(\frac{s^*(v_i, \Delta v_i)}{s_i} \right)^2\bigg),
\end{cases}
\]

where $s^*(v_i,\Delta v_i) = s_0 + v_i T + \frac{v_i\Delta v_i}{2\sqrt{ab}}$, given model parameters $v_0, s_0, T, a, b$. Here, $v_{0}$ is a maximal target speed that could be chosen as $30 m.s^{-1}$, $s_{0}$ is a minimum distance that could typically be chosen as $2m$, $T$ is a characteristic time (or safe timeheadway) that could be chosen as $1.5s$, $a$ a maximal acceleration and $b$ a maximal comfortable deceleration that could be chosen respectively as $1m.s^{-2}$ and $2m.s^{-2}$.\\

For the Bando model, introduced in \cite{bando1995dynamical}, the longitudinal dynamics for a lane are written:

\[
\begin{cases}
\dot{x}_i = v_i,\\
\dot{v}_i = \alpha \bigg(V(\Delta x_i)-v_i\bigg),
\end{cases}
\]

where $V(\cdot)$ is the optimal velocity function that depends on the space headway, $\Delta x_i=x_{i+1}-x_i$, in front of the $i^{\text{th}}$ vehicle (see \eqref{defB}).

For the FTL model, introduced in \cite{gazis1961nonlinear} (see also \cite[Eq. (19)-(20)]{helbing2001traffic}), the longitudinal dynamics for a lane are written:

\[
\begin{cases}
\dot{x}_i = v_i,\\
\dot v_{i} = \beta\frac{v_{i+1}-v_{i}}{(x_{i+1}-x_{i}-l_{v})^{2}}
\end{cases}
\]

Not all models recreate stop and go traffic waves. Regarding this phenomena, the IDM or a combination of the Bando model and FTL, so called ``Bando-FTL model'', are usually used \cite{treiber1999explanation,delle2019feedback,cui2017stabilizing}, although a modified version of the original Bando model including some delays (see \cite{bando1998analysis}) could also be used. The well-posedness of the Bando-FTL model and its delayed version were studied in \cite{GongKeimer2022}.
For the Bando-FTL, the longitudinal dynamics for a lane are written:

\begin{equation}
\label{Bando}
\begin{cases}
\dot{x}_i = v_i,\\
\dot v_{i} = \alpha(V(x_{i+1}-x_{i})-v_{i})+\beta\frac{v_{i+1}-v_{i}}{(x_{i+1}-x_{i}-l_{v})^{2}},
\end{cases}
\end{equation}

where $v_{i}$ is the velocity of the $i^{\text{th}}$ car and $x_{i}$ is its location. The constant $\alpha$ is the weight for the Bando model and $\beta$ is the weight of the Follow-the-leader model. $V$ is still the optimal velocity function given by \cite{bando1995dynamical,delle2019feedback}

\begin{equation}
V(x)=V_{\max}\frac{\tanh(\frac{x-l_{v}}{d_{0}}-2)+\tanh(2)}{1+\tanh(2)}.
\label{defB}
\end{equation}

where $l_v$ is the length of the car, and $d_0$ is the minimal distance for the optimal velocity model (see Table \ref{table1}). The Bando-FTL model is primarily studied in this paper. This model has been used in the past, for instance in \cite{delle2019feedback,cui2017stabilizing}, and has several advantages : 
\begin{itemize}
    \item the FTL model represents the competing dynamics between drivers and deals with the safety issues by applying a large braking value when a vehicle is too close to the leading vehicle. This portion is at the origin of the stop-and-go waves.
    \item the Bando model enables realistic uniform flow steady-states: for the density of cars on the road, there is a unique uniform flow equilibrium $(h,v^{*}(h))$, where $h$ is the equilibrium headway and $v^{*}(h)$ is the equilibrium speed, which decreases with $h$. Different optimal velocity profiles may be used, but then car passing is difficult to prevent and 
unrealistic large acceleration may be generated, see \cite{1998,articlexq}.
    \item Because the FTL portion already incorporates a safety criteria, given reasonable initial conditions, the Bando-FTL model usually does not require an additional fail/safe condition. The model is also generally more robust than the widely used Intelligent Driver's Model (IDM) \cite{treiber2000congested}.
\end{itemize}
Moreover, if we consider $N$ vehicles on a multilane ring-road with only human-driven vehicles, the traffic is represented by the variables $(x_{i}^{j},v_{i}^{j})_{i\in \{1,...,n^{j}\},\; j\in J}$, where $j$ is the lane number, $n^{j}$ is the number of vehicles in lane $j$. The dynamics of these variables still follows the Bando-FTL model \eqref{Bando} with the notations $v^{j}_{n^{j}+1}:= v_{1}^{j}$ and $x^{j}_{n^{j}+1}:=x_{1}^{j}$ to take into account the ring-road geography.

\textcolor{black}{To take into account physical limitations of real cars, we also cap the acceleration to $2.5$ $m{\cdot}s^{-2}$ and the deceleration to $4$ $m{\cdot}s^{-2}$.}
It has been shown that this model produces stop-and-go waves in lane $j$
if \cite{cui2017stabilizing}

\begin{equation}
\label{unstable}
\frac{\alpha}{2}+\frac{L^{2}\beta}{(n^{j})^{2}}< V'\left(\frac{n^{j}}{L}\right).
\end{equation}

\vspace{\baselineskip}
where $V'$ is the derivative of function $V(x)$ in equation \eqref{defB}, and $L$ is the length of the road.

\vspace{\baselineskip}
{\itshape \textcolor{orange}{b}) Lateral dynamics and lane-change mechanism}\\
Regarding the Bando-FTL model, we include a lane changing mechanism suggested by Treiber et al. in \cite{treiber2009modeling}.
Several models of lane changing dynamics have been explored \cite{zheng2014recent}, such as 1. Gipps-type lane changing \cite{gipps1986model,yang1996microscopic,hidas2002modelling}, and 2. Utility theory based lane changing \cite{ahmed1996models,toledo2003modeling}.
In Gipps-type lane changing, the driver's behavior is governed by maintaining a desired speed and being in the correct lane for an intended maneuver. These types of lane changes depend on parameters corresponding to an incentive and an acceptable level of risk for a collision, where some differentiate between cooperative and forced lane changes. Characteristics distinguishing utility theory based lane changing are a hierarchical decision-making process, desirability versus necessity, and the consideration of multiple driver types (driver behavior heterogeneity).

A regular vehicle changes lane if and only if
\begin{itemize}
    \item It is safe to do so: changing lane does not imply a huge braking for the vehicle behind.
    \item It has an acceleration incentive: the expected acceleration after changing lane is higher than the expected acceleration from not changing lane.
    \item A certain amount of time has passed from the time of the vehicle's last lane change to the current time. We refer to this as the lane change ``cooldown time.''
\end{itemize}
In mathematical form, if we denote $i$ as the vehicle changing lane and $j$ as the potential new lane, we have

\begin{equation}
\begin{split}
&\tilde a_{i}^{j}>a_{i}+\Delta_{I}\;\;\text{(incentive)},\\
&\tilde a_{i}^{j}>-\Delta_{s}, \;\; \tilde a_{\text{fol}}^{j}(i)>-\Delta_{s}\;\;\text{(safety)},\\
&t > t_0+\tau  \;\;\text{(cooldown time)}.\\
\end{split}
\label{safety}
\end{equation}

Here $a_{i}$ is the acceleration of the vehicle changing lane in the original lane, $\tilde a_{i}$ is the expected acceleration in the new lane, $\tilde a_{\text{fol}}^{j}(i)$ is the expected acceleration of its follower in the new lane, $\Delta_{I}$ is a constant representing the threshold incentive and $\Delta_{s}$ represents the threshold safety. For the cooldown time equation, $t_0$ represents the last time a lane change occurred for the considered vehicle, and shows that the time of the next lane change should be greater by a threshold value $\tau$.

There are two main advantages to using acceleration, instead of speed, to model lane-change: $(1)$ the lane-change decision-making process is dramatically simplified; $(2)$ one can readily calculate accelerations with an underlying microscopic longitudinal traffic model, see \cite{zheng2014recent}.
We also point out that the lane-change mechanisms lead to discrete dynamics of the vehicles. The presence of both continuous dynamics and discrete dynamics of vehicles motivate us to consider a hybrid system, see \cite{654885,garavello2005hybrid,4806347,piccoli1998hybrid,Tomlin_1998}. 

A natural question is to wonder about the influence of the lane-changing mechanisms on the stability of the system and whether such a model reduces traffic instabilities when adding the lane-changing mechanisms, or on the contrary, whether it produces even stronger traffic instabilities.
We show in the next section that the model heavily depends on the parameters of the lane-changing mechanisms. Besides this, the possible behaviors are extensive. As the lanes are coupled, there are scenarios where one lane can produce instabilities such as stop-and-go waves while another lane does not.

\section{Strong influence of the lane-changing parameters on the traffic behavior}
\label{sec:2}
\begin{figure}[H]
\centering
\includegraphics[width =0.7\textwidth]{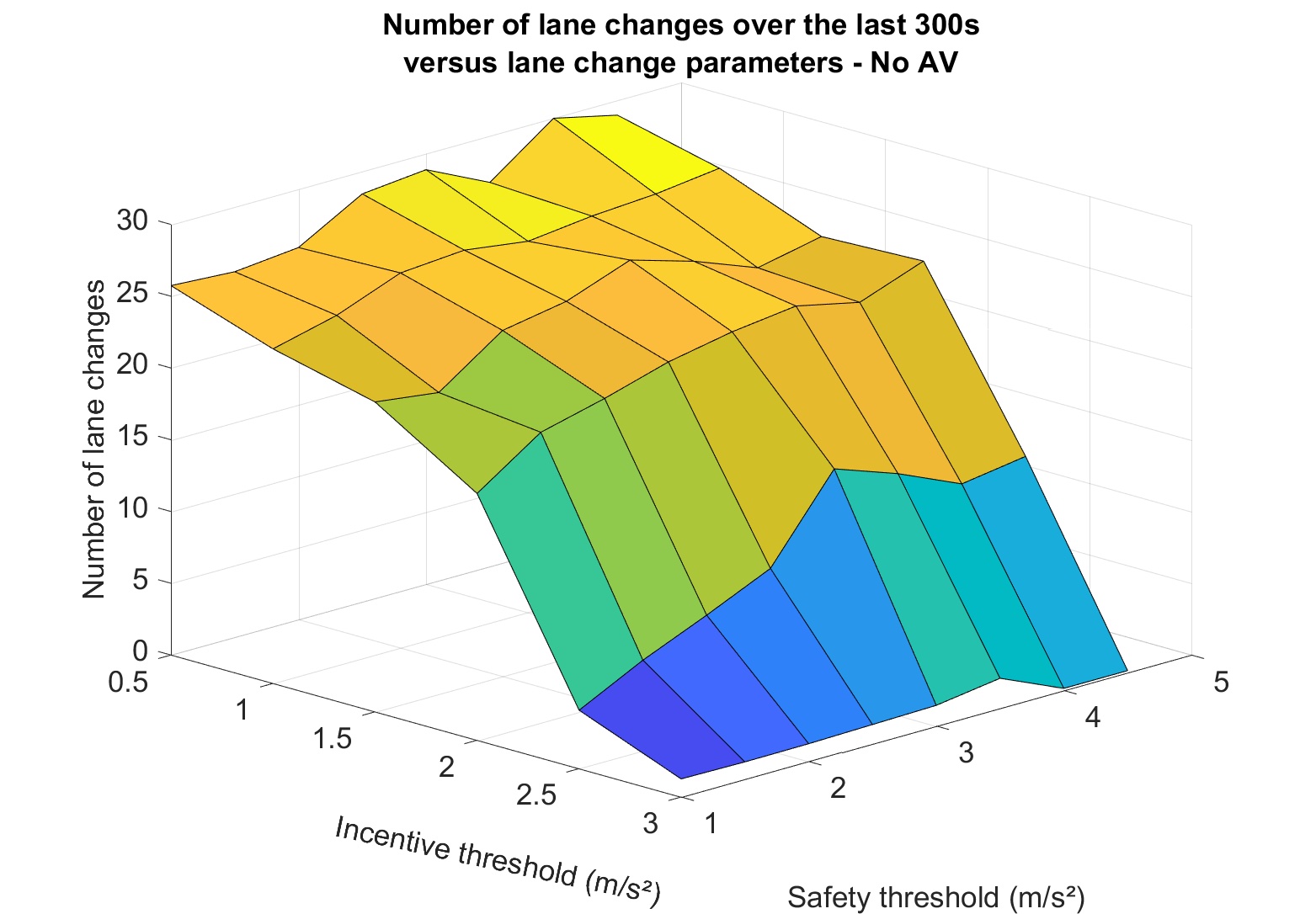}
\caption{Number of lane changes without control in the system given different threshold values for incentive and safety
\label{fig0}}
\end{figure}

In this section we study the effect of the threshold parameters $\Delta_{I}$ and $\Delta_{s}$ and we show that different values of these parameters can lead to radically different behaviors for the traffic flow. To illustrate this phenomena, we fix a given initial condition where all lanes have the same number of cars (in this case 24 cars for the middle lane of length 240m, see Table \ref{table1} for a summary of the parameters used), and all the cars are initially located within $1m$ from their steady-state location (the steady-state location corresponds to a uniform spacing). Then we perform traffic simulations over $1000s$ with $\Delta_{I}$ ranging from 0.6 to 3 $m{\cdot}s
^{-2}$ and $\Delta_{s}$ ranging from 0.5 to 5 $m{\cdot}s
^{-2}$. Note that in our model, the maximum acceleration allowed by the car is $2.5 m{\cdot}s^{-2}$ and the maximum deceleration allowed by the car is $4 m{\cdot}s^{-2}$. The explanation for the choice of the range on $\Delta_{I}$ and $\Delta_{s}$ is as follows: requiring an incentive $\Delta_{I}$ of $3m{\cdot}s^{-2}$ to change lane means that you can only change lane when your lane is decelerating and you can accelerate strongly in the neighboring lane, whereas a safety threshold $\Delta_{s}$ of $5m{\cdot}s
^{-2}$ means that we do not require any safety since the vehicles cannot brake more strongly anyway. We expect that the higher the request on the incentive is, the lower the number of lane changes. Similarly the higher the security threshold (hence the lower security required), the higher the number of lane changes. These expectations were confirmed in Figure \ref{fig0}, where we plot the number of lane changes over the total length of the simulation for each combination of parameters $\Delta_{I}$ and $\Delta_{s}$.

\begin{figure}[H]
\centering
\includegraphics[width =0.48\textwidth]{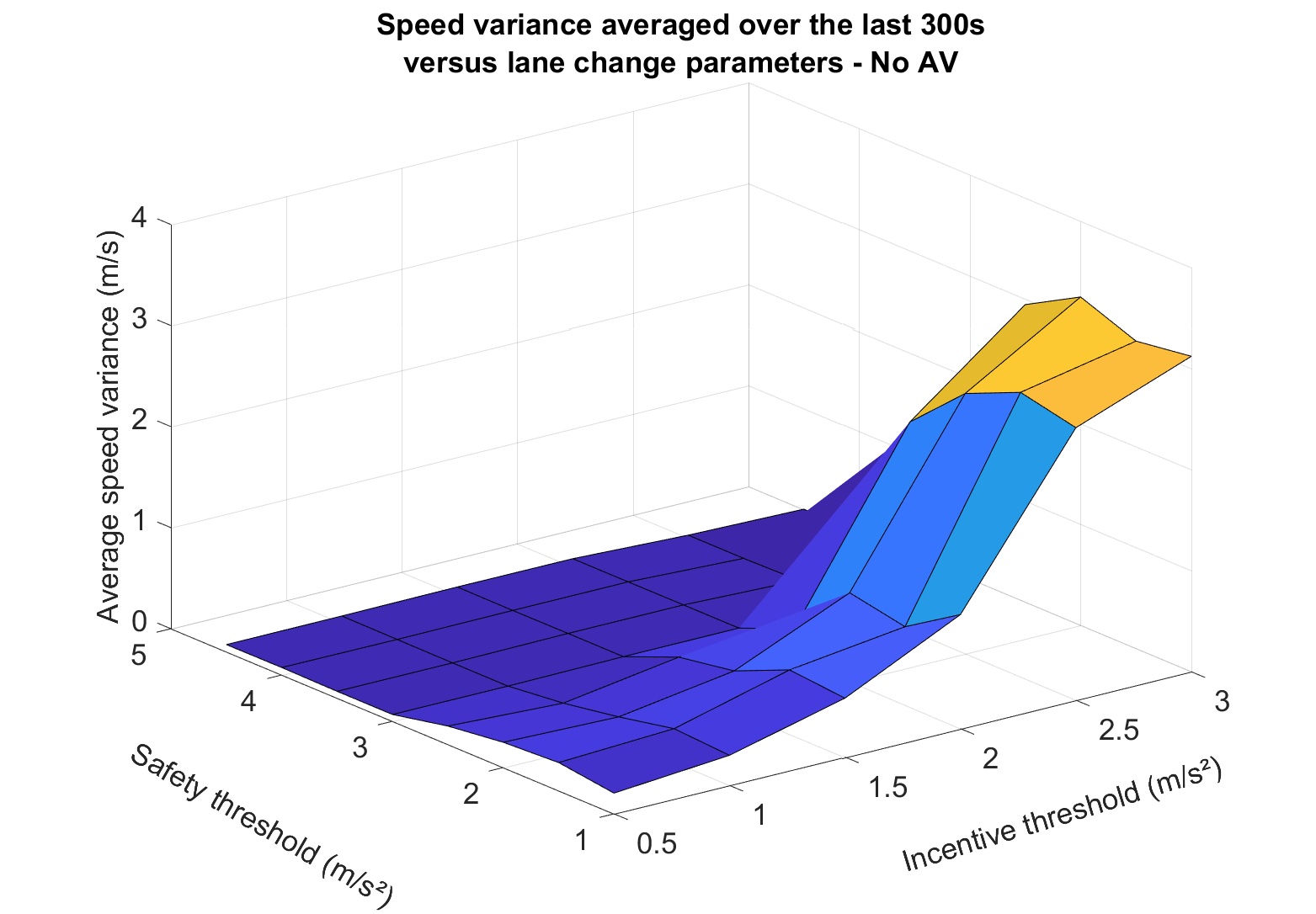}
\includegraphics[width =0.48\textwidth]{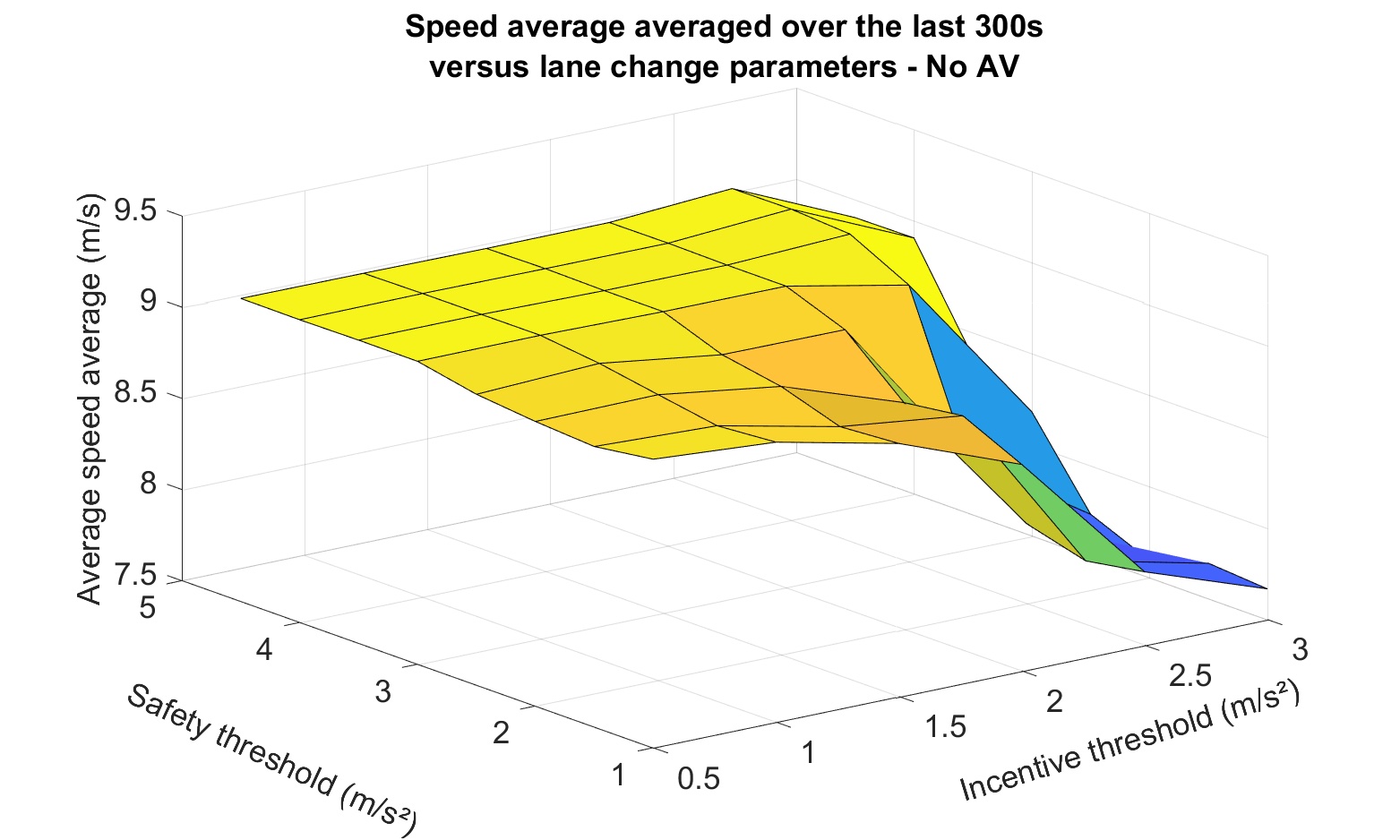}
\caption{Speed variance and average speed of the system without control given different threshold values for incentive and safety
\label{fig1}}
\end{figure}

In Figure \ref{fig1}, for each simulation, we compute the speed variance for each lane and at each time-step, and find the average of the value over the \textcolor{black}{last 300s} and across the three lanes. \textcolor{black}{We then average this over 100 simulations with random initial conditions} We see that in some cases the speed variance is close to 0, which suggests that the system reached equilibrium. In other cases the speed variance has a high value, suggesting that the system still \textcolor{black}{undergoes} some stop-and-go waves. 

\begin{figure}[h!]
\centering
\includegraphics[width =0.45\textwidth]{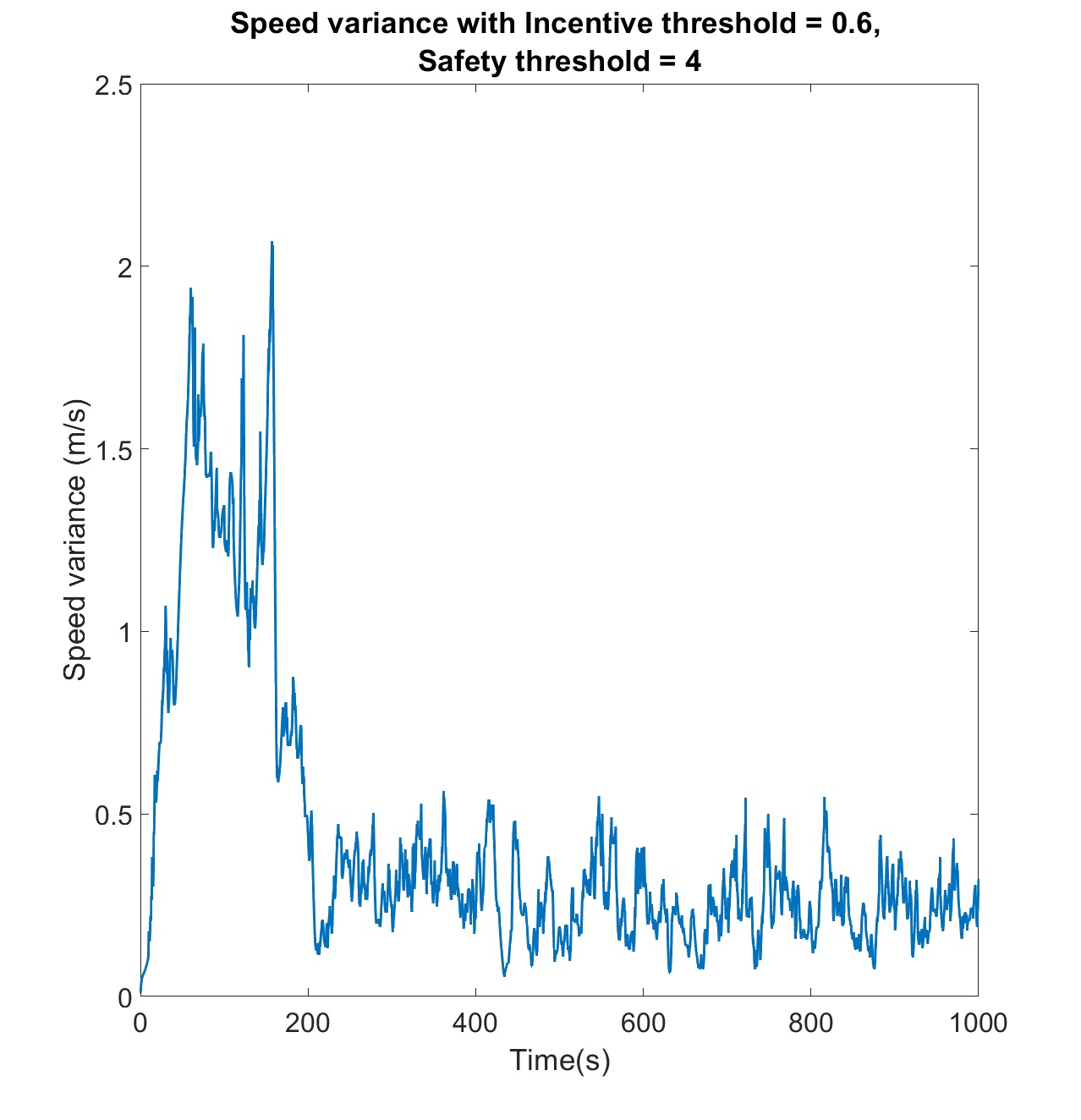}
\includegraphics[width =0.45\textwidth]{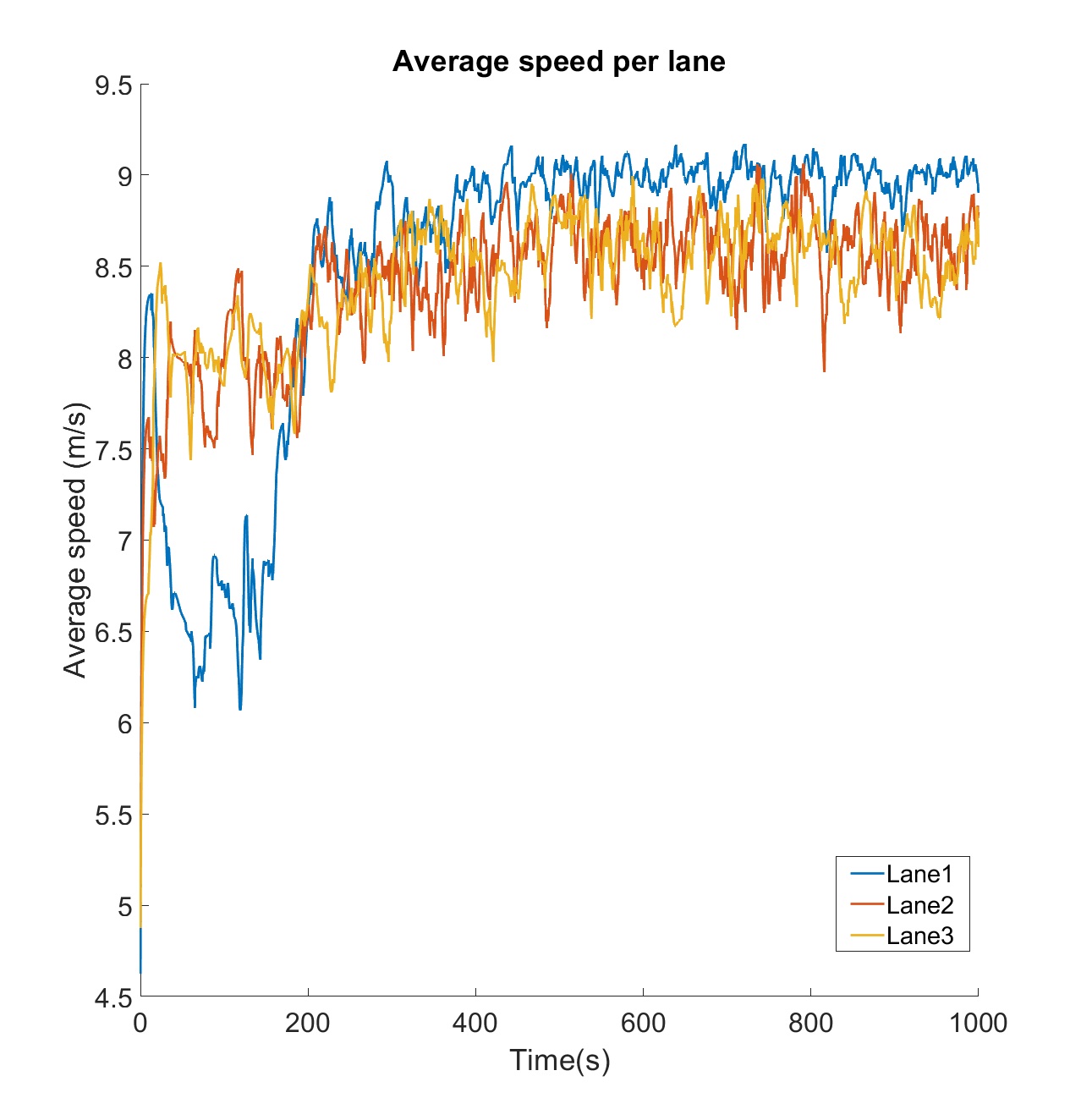}
\includegraphics[width =0.45\textwidth]{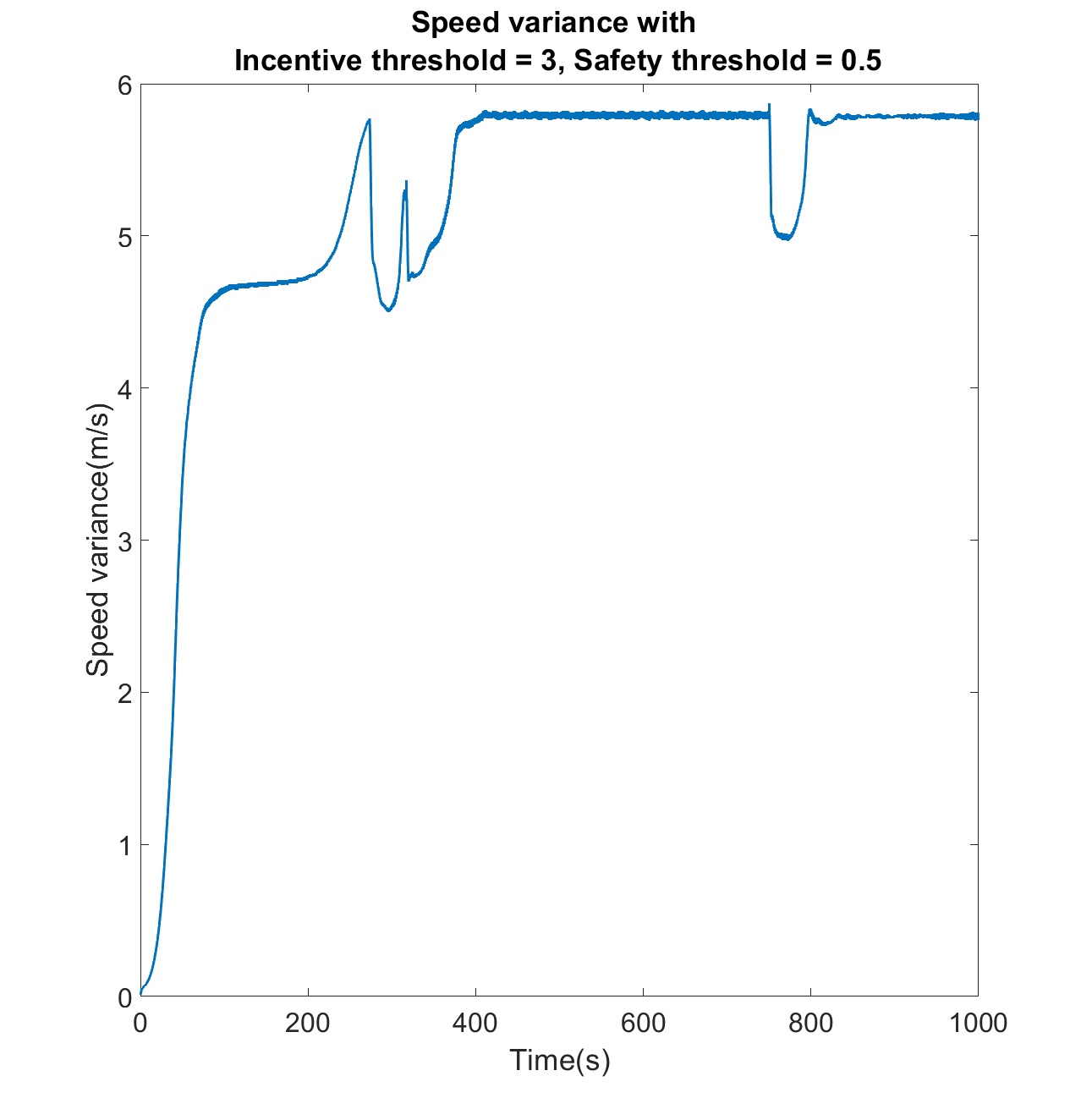}
\includegraphics[width =0.45\textwidth]{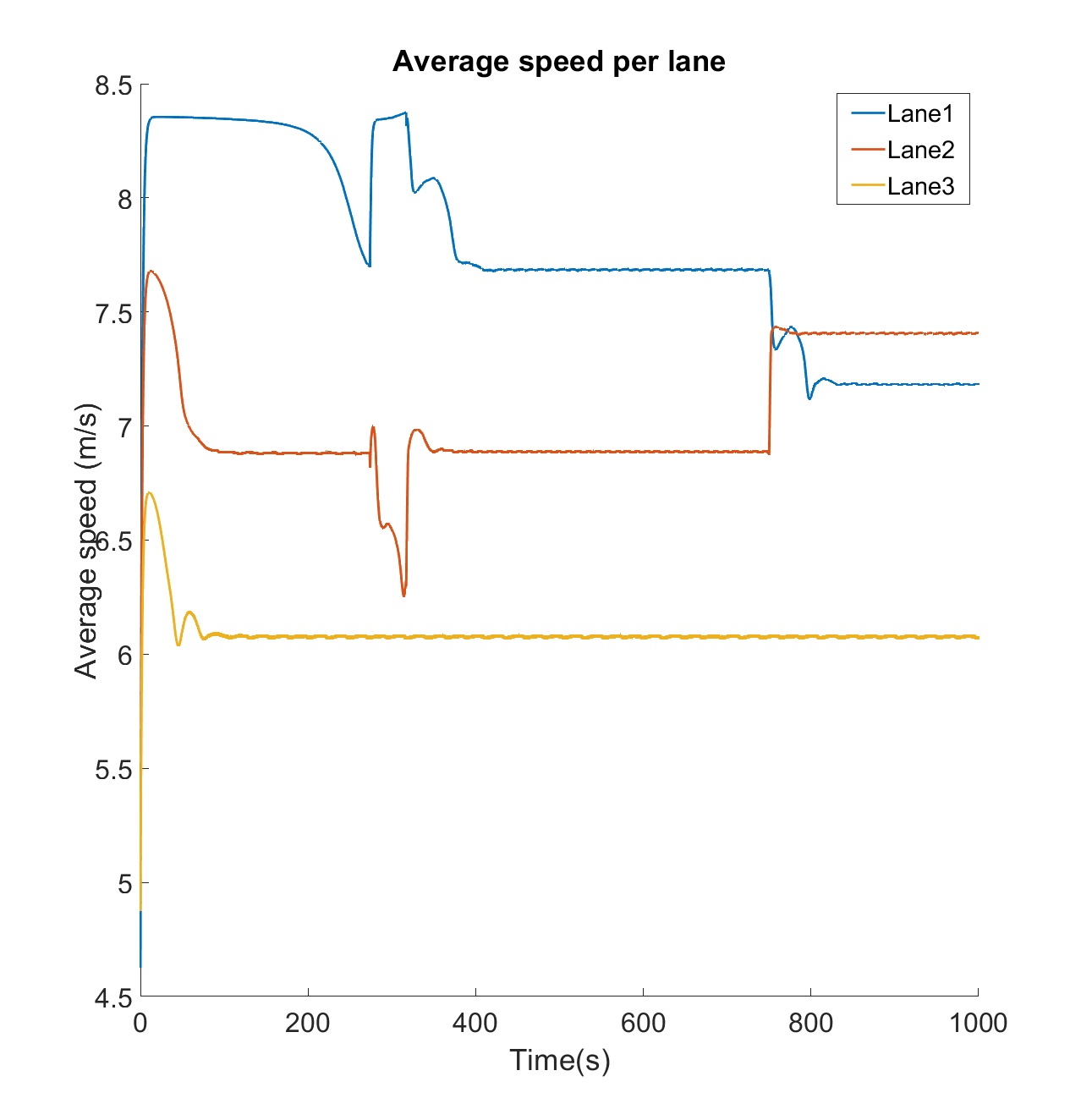}
\caption{Speed variance and average speed over time for different threshold values. {\small Up left:speed variance with incentive threshold $=0.6m.s^{-2}$ and safety threshold $=4m.s^{-2}$, 
Up right: the average speed per lane with the same incentive and safety thresholds.
Down left: speed variance with incentive threshold $=3m.s^{-2}$ and safety threshold $=0.5m.s^{-2}$, Down right: the average speed per lane with the same incentive and safety thresholds.}
\label{fig1_8}}
\end{figure}

These speculations seems to be clear when looking at the instantaneous speed variance over time for Fig \ref{fig1_8} with parameter values $\Delta_{s}= 4$ $m{\cdot}s^{-2}$, $\Delta_{I}=0.6$ $m{\cdot}s^{-2}$ and $\Delta_{s}=0.5$ $m{\cdot}s^{-2}$ $\Delta_{I}=3$ $m{\cdot}s^{-2}$ respectively. We see that for  $\Delta_{s}=4$ $m{\cdot}s^{-2}$ and $\Delta_{I}=0.6$ $m{\cdot}s^{-2}$, the system approaches a uniform flow after $200$ s, while for $\Delta_{s}=0.5$ $m{\cdot}s^{-2}$ and $\Delta_{I}=3$ $m{\cdot}s^{-2}$, it does not and traffic instabilities persist in the system. Comparing this to Figure \ref{fig0}, one can note that in the area with a very small incentive threshold and a very large safety threshold, there are (logically) a large number of lane-changes but no apparent traffic instability, a safety variance close to $0$ and a high average velocity. This is something that can seem counter-intuitive. In fact in this situation lane-changes happens each time there is a slight difference of speed in the lane, and nearly no matter unsafe it is. This seems to result in an homogenization of the traffic. Of course this could contradict real-life experience because in this extreme case the lane-changes are very non-human (they are in particular extremely dangerous).
\\

\section{Using autonomous vehicles to smooth traffic instabilities.}
\label{sec:3}

Traffic flow is very particular in that a single individual can have a global effect on the entire dynamic of the flow. This is found in both micro and macro models and can be understood from a simple example: a single individual can be a bottleneck and thus influence the traffic across the entire system. Given this, the section serves to investigate the following: is it possible to dissipate and prevent traffic instabilities by simply adding a single 
AV that follows a prescribed acceleration? And if so, what prescribed acceleration should be given to these cars in order to smooth traffic efficiently? \\

When adding an AV to the system, the equations are modified as follows: the lane of the AV is denoted by $j$ and the car's number is denoted by $0$. From this, we have

\begin{equation}
\begin{split}
\dot x_{0}(t) &= v_{0}(t),\\
\dot v_{0}(t) &= u(t),
\end{split}
\end{equation}

where $u$ is a control law that can be chosen.
Using AVs to smooth traffic flow has already been studied in both theory and experiments in a single lane context \cite{delle2019feedback}. In particular, from \cite{delle2019feedback}, the author uses two very simple controllers, one proportional and one slightly proportional integral controller. From the theoretical analysis and experiments, the author demonstrated the efficiency of such simple controllers to smooth stop-and-go waves in a single lane ring-road, with a reduction of fuel consumption of up to $40\%$. However, when the traffic is multi-lane, the problem becomes more difficult for several reasons:
\begin{itemize}
    \item The lane-changes add complexity to the dynamics of the system and impacts the stability of the stop-and-go waves, potentially making them harder to smooth.
    \item The AV only belongs to one lane but can dissipate and prevent waves on all three lanes. Hence, on two lanes, the lane-changes are represented by the coupling between waves.
    \item The model is very sensitive to errors from the parameters, as shown in the previous section, and these errors could lead to simulations that are far from the ground truth.
\end{itemize}

The controller we use in our setting is a proportional controller where the ideal command is given as follows:

\begin{equation}
\begin{split}
u(t) &= -k\left(v_{0}-v_{\text{target}}\right),\\
v_{\text{target}} &= v^{*}\left(\frac{n^{j}+l_{v}}{L^{j}}\right),
\end{split}
\end{equation}

 where $l_{v}$ is the average length of a car, $k$ is a constant design parameter, $L^{j}$ is the length of lane $j$ and $n^{j}$ is the total number of cars in the j-th lane. Recall that $v^{*}(h)$ is the steady-state speed of the system corresponding to the steady-state headway $h$. $v_{target}$ is the speed of the uniform flow steady-state we would like to reach. This ideal command does not take into account to prevent the AV from crashing into another car. To tackle this, one could add a safety mechanism where the AV would brake if it is too close to its leader. However, even with a safety mechanism, the AV can still get stuck in stop-and-go waves. This is because $v_{\text{target}}$ would be too high compared to the current velocity of the cars in front of the AV. The AV would then try to increase its speed until it is too close to the vehicle in front, then it would brake, and then increase its speed again, thus maintaining a stop and go wave. Due to this, we add the following features to our control: 
 \begin{itemize}
     \item (quasi-stationary steady-state strategy) \textcolor{black}{As mentioned, we are trying to make the AV not get stuck in a stop and go wave as it tries to reach an ideal steady-state speed that is higher than the speed of its leader. To deal with this, we start by stabilizing a smaller speed, and then we slowly raise the stabilizing speed to the ideal steady state speed. In control literature, this is referred to as following a continuous path of a steady-state. This is only possible because adding the AV allows the number of possible steady-states to go from a single steady-state to a continuous range. In mathematical terms, the control law becomes
     \begin{equation}
        \begin{split}
u(t)&=-k(v_{0}-\bar v_{d}(t)),
\end{split}
     \end{equation}
     where $v_{d}$ is given by
     \begin{equation}
\left\{\begin{split}
v_{d}(t)=&v_{\min}+(\bar v^{*}(h^{*})-v_{\min})\frac{t}{t_{tr}},\text{  for }t\in [0,t_{tr}],\\
v_{d}(t)=&\bar v^{*}(h^{*}),\text{  for }t\geq t_{tr}\\
\end{split}\right.
\end{equation}
$t_{tr}$ is the time of transition and $v^{*}(h^{*})$ is the ideal steady-state speed.}
\\
     \item (safety mechanism) \textcolor{black}{When the AV starts to get close to its leading vehicle we change the target speed to the speed of the leading vehicle for safety.}
 \end{itemize}
\vspace{\baselineskip}

{\itshape a) Lateral controller}\\
In a multilane framework, another interesting means of control for the AV is having the ability to change lanes, and this is referred to as a lateral controller. Given the results from \cite{delle2019feedback}, traffic can be stabilized with one AV per lane in the case that the AVs cannot change lanes. However, if AVs can change lanes and have good lateral controllers, then traffic can be stabilized in multilane ring-roads with potentially even a single AV. \textcolor{black}{Our lateral controller is the following: the AV changes lane if and only if
\begin{itemize}
\item the safety conditions (\ref{safety}) are satisfied (just like for a regular vehicle).
\item the speed variance in another lane averaged on the last $t_{1}$ seconds, is higher than the speed variance in the AV's lane, also averaged on the last $t_{1}$ seconds. This difference has to be larger than a threshold (noted $c_{1}$ in Table \ref{table1}).
\item the AV has not been changing lanes in the last $t_{2}$ seconds.
\end{itemize}
}
\textcolor{black}{
We denote the AV's lane by $j_{0}$, and the last time the AV changed lane as $t_{0}$ ($t_{0} =0$ if the AV never changed lane). From this, we have
\begin{itemize}
    \item $t>t_{1}$ and there exists $j\in\{1,...,3\}\setminus\{j_{0}\}$ such that
\begin{equation}
\begin{split}
\int_{t-t_{1}}^{t}&\frac{1}{N_{j}}\sum\limits_{i=1}^{N_{j}}(v_{i}^{j})^{2}(s)-\frac{1}{N_{j}^{2}}\left(\sum\limits_{i=1}^{N_{j}}v_{i}^{j}(s)\right)^{2} ds\\
&>c_{1}+\int_{t-t_{1}}^{t}\frac{1}{N_{j_{0}}}\sum\limits_{i=1}^{N_{j_{0}}}(v_{i}^{j_{0}})^{2}(s)-\frac{1}{N_{j_{0}}^{2}}\left(\sum\limits_{i=1}^{N_{j_{0}}}v_{i}^{j_{0}}(s)\right)^{2} ds.
\end{split}
\end{equation}
\item \textcolor{black}{$t>t_{2}+t_{0}$}.
\item the safety condition \eqref{safety} is satisfied with $i=0$ and $j=j_{0}$.
\end{itemize}
The main difference between the regular vehicles and the AV is that the incentive for the AV is to go in the lane with the highest speed variance. This is different to an incentive that is based on acceleration. Averaging and threshold values are included to account for the stochastic nature of the measurements, and to avoid the AV changing lanes constantly, which could destabilize the system. Note that this lateral controller assumes a global knowledge of the state of the system, which might be a limitation in practice. Nevertheless, with V2V connectivity coupled to the fact that this information is only accessed at a reduced frequency (cooling time $t_{2}$ and evaluation time $t_{1}$ are set as 10$s$) we can hope of reconstructing the knowledge of the speed variance with an observer. In lack of V2V, side sensors and history could also allow to construct at least a rough estimate of this quantity.
}\\

\begin{table}
\begin{center}
 \begin{tabular}{||c c c||} 
 \hline
 Parameter & Value & Description \\ [0.5ex] 
 \hline\hline
 N & 24 & number of vehicles per lane \\ 
 \hline
 J & 3 & number of lanes \\ 
 \hline
 $l_{v}$ & 4.5 & length of a car [$m$] \\ 
 \hline
 $d_{0}$ & 2.5 & minimal distance for optimal velocity model \\ 
 \hline
 $\beta$ & 20 & weight FTL \\ 
 \hline
 $\alpha$  & 0.5 & weight OV \\ 
 \hline
 $dt$ & 0.02 & timestep size for the simulation \\ 
 \hline
 $t_{f}$ & 1000 & final time of the simulation [$s$] \\ 
 \hline
 max-dec & 4 & maximum deceleration [$m{\cdot}s^{-2}$] \\ 
 \hline
 max-acc & 2.5 & maximum acceleration [$m{\cdot}s^{-2}$] \\ 
 \hline
 iter-lc & 50 & iteration for lane changing, dependent on dt \\ 
 \hline
 $\tau$ & 5 & cool down duration after lane change [s] \\ 
 \hline
 k & 1 & constant in control law AV \\ 
 \hline
 $c_{1}$ & 0.5 & speed variance threshold for AV changing lane \\   
 \hline
 $t_{1}$ & 10 & time to average speed variance for AV changing lane [s] \\ 
 \hline
 $t_{2}$ & 10 & AV cool down duration after lane change [s] \\
 \hline
\end{tabular}
\end{center}
\caption{Parameters used for the simulations \label{table1}}
\end{table}

\begin{figure}[H]
\centering
\includegraphics[width =0.48\textwidth]{speedvar_no_av}
\includegraphics[width =0.48\textwidth]{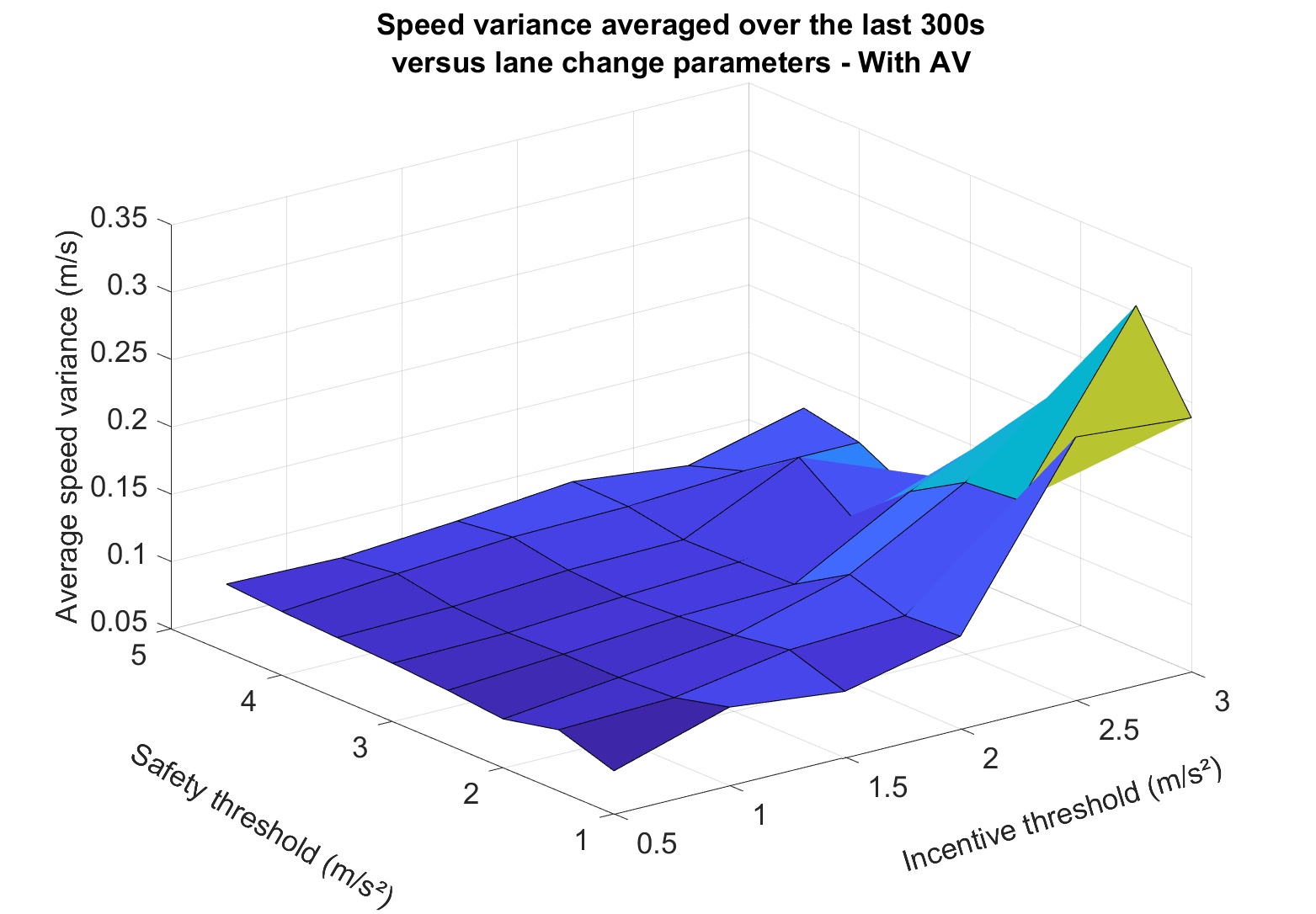}
\caption{Speed variance for different safety and incentive thresholds. Left: without control, Right: with control.
\label{fig_no_control_last_sec}}
\end{figure}

{\itshape b) Results}\\
In this section we show that, similar simple controllers not only manage to smooth traffic instabilities in particular stop-and-go waves in a multi-lane setting, but also hold a large range of parameters $\Delta_{I}$ and $\Delta_{s}$. We run two batches of simulations with a fixed initial condition \textcolor{black}{with random perturbations as in Fig. \ref{fig1}} and different parameters of $\Delta_{I}$ and $\Delta_{s}$. \textcolor{black}{For each set of parameters we run 100 simulations and average the results over the 100 simulations and the 300 last seconds.} The first batch in the experiment is similar to the previous section in that there are no AVs. In the second batch of the experiment, we turn an AV on. The AV is initially in the middle lane. In Figure \ref{fig_no_control_last_sec}, for simulations with and without the AV respectively, we represent the speed variance averaged over \textcolor{black}{the 300 last seconds} and the three lanes, for each pair of parameters $\Delta_{I}$ and $\Delta_{s}$. 
\textcolor{black}{We} see a significant difference between the speed variance of the system without control (left) and without control (right): the speed variance when adding the AV is always below $0.3 m.s^{-1}$, namely \textcolor{black}{ten} times smaller that the case without control.

\begin{figure}[H]
\centering
\includegraphics[width =0.4\textwidth]{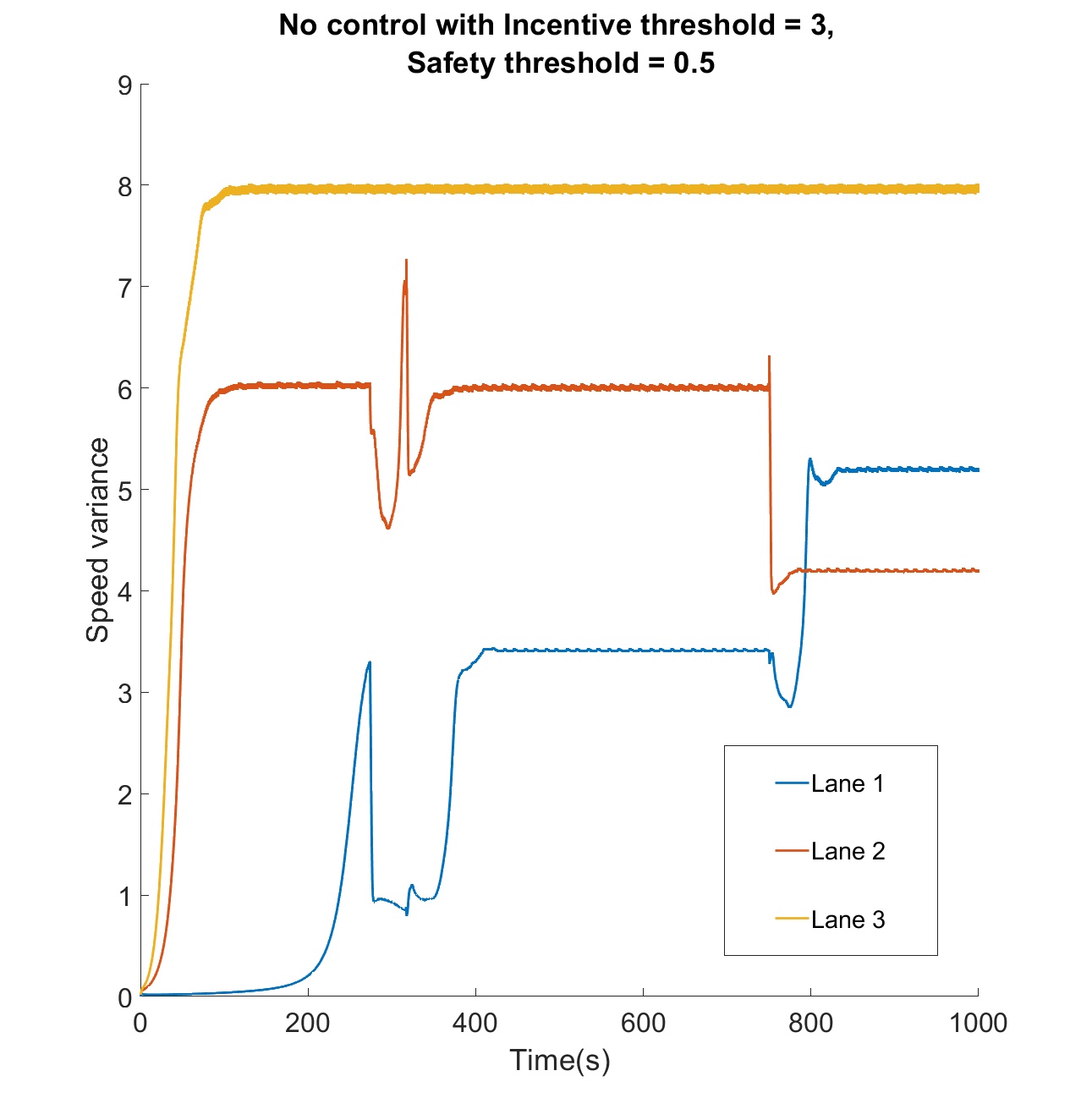}
\includegraphics[width =0.4\textwidth]{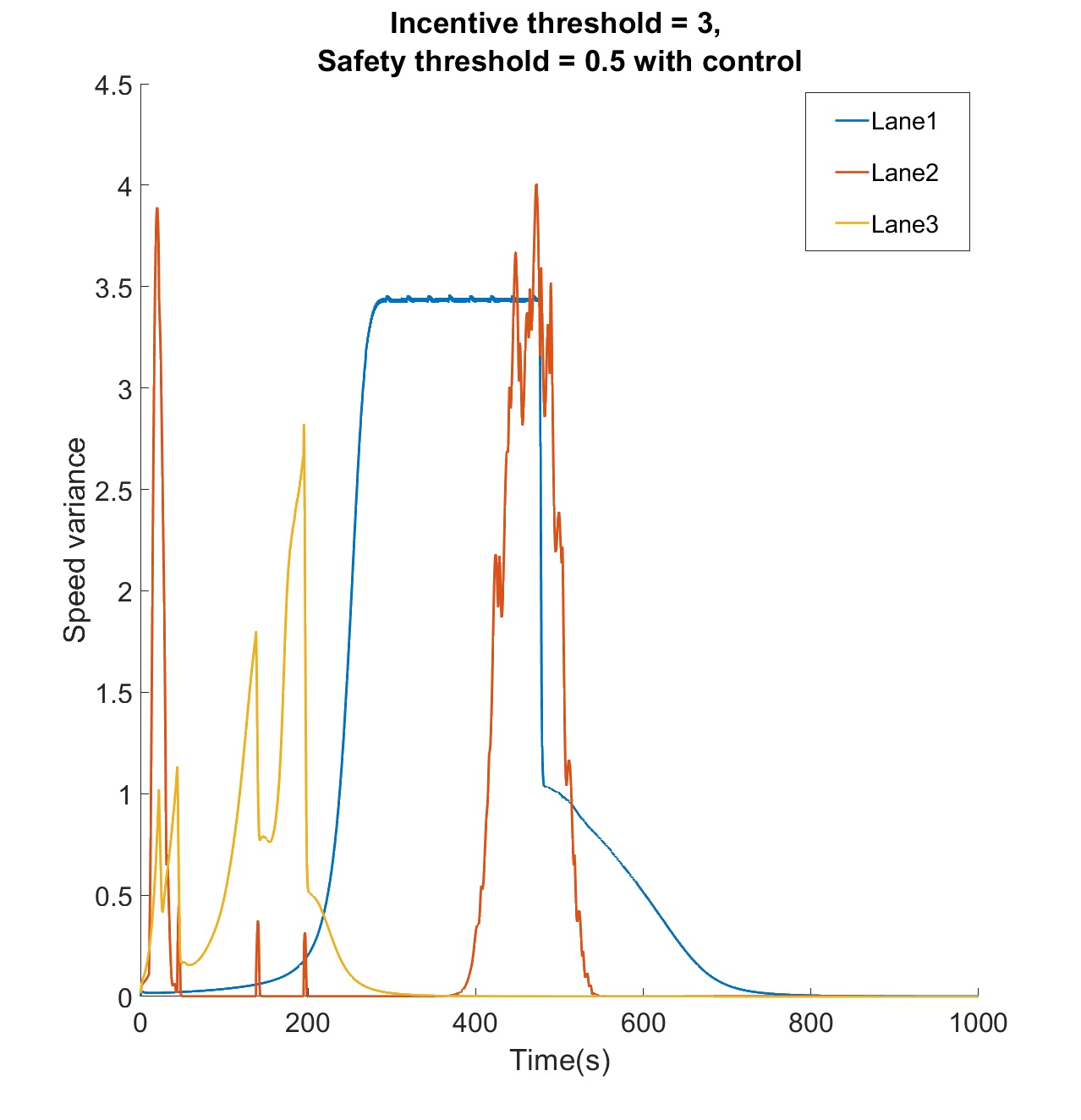}
\caption{Speed variance over time for different threshold parameters. Left: without control, Right:  with control.
\label{speed_variance_three_lanes_3_5}}
\end{figure}

Note that the reduction of speed variance and dissipation of waves is effective over all the range of parameters $\Delta_{I}$ and $\Delta_{s}$. Moreover, there is a single AV in this simulation, therefore the penetration rate (fraction of AV in the total traffic) is below $2\%$. To illustrate what is going on, we represent in Figure \ref{speed_variance_three_lanes_3_5} the speed variance over time in all three lanes, where $\Delta_{I}=3$ $m{\cdot}s
^{-2}$ and $\Delta_{s}=0.5$ $m{\cdot}s
^{-2}$, both with and without a controller. As expected, we see that the AV is stabilizing mostly one lane that reaches uniform flow, but despite the very weak coupling of the lanes (due to the very small number of lane changes), it is still enough to roughly dissipate the waves that form in the other lanes. On the other hand, when there is no AV, the speed variance remains high. Note that the y-axis in the figure with the control only goes to 4.5 $m{\cdot}s
^{-1}$ while the axis of the figure without the control goes to 9 $m{\cdot}s^{-1}$.

\section{Collaborative driving}
\label{sec:4}
Another promising approach that could highly impact traffic inefficiencies, including dissipation
of traffic instabilities is the collaborative driving behavior of the drivers.
\textcolor{black}{Collaborative driving (CD), also combined with autonomy \cite{9138645,5326205}, is an important emerging aspect of Intelligent Transportation Systems (ITS). CD is often times based on communication with various possible approaches proposed in the literature \cite{1398942,4591315,LM16,PETRILLO2018372,6459954,Sannier02},
and needs to take into account human behavior
\cite{KLN20,8556397}.}\\

\textcolor{black}{Here, we take the simple approach of assuming that a fraction of the human drivers is instructed to target specific preferred speed, while keeping a smooth and safe driving. This represents 
an offline centralized control mechanism, with decentralized human actuators. The target speed may be communicated daily or for times of day.
More precisely,
denoting $S$ the number of vehicles in collaborative driving and $I$ the rest of the vehicles in the lane, we have
\begin{equation}
\left\{\begin{split}
\dot{x}_i &= v_i, \quad i=1, \dots, n,\\
\dot v_{i} &= \alpha_{i}(V(x_{i+1}-x_{i})-v_{i})+\beta_{i}\frac{v_{i+1}-v_{i}}{(x_{i+1}-x_{i}-l_{v})^{2}},
\end{split}\right.
\end{equation}
where $n$ is the number of vehicles, $\alpha_{i}= \alpha$ and $\beta_{i}=\beta$ if $i\in I$,  $\alpha_{i}=\alpha_{S}$ and $\beta_{i}=\beta_{S}$ if $i\in S$, with $\alpha$ and $\beta$ such that \eqref{unstable} holds and $\alpha_{S}$ and $\beta_{S}$ satisfy the opposite inequality 
\begin{equation}
    \label{stable}
    \frac{\alpha}{2}+\frac{L^{2}\beta}{(n)^{2}}> V'\left(\frac{n}{L}\right).
\end{equation}
We present here numerical simulations suggesting that such a collaborative behavior allows to recover some stability of the flow and decreases speed variance and car accelerations, and hence energy consumption. However, while promising, these results also suggest that AVs are much more efficient at recovering stabilities of the flow with a very small penetration rate.} \\

\begin{figure}[h!]
    \centering
    \includegraphics[width=0.98\textwidth]{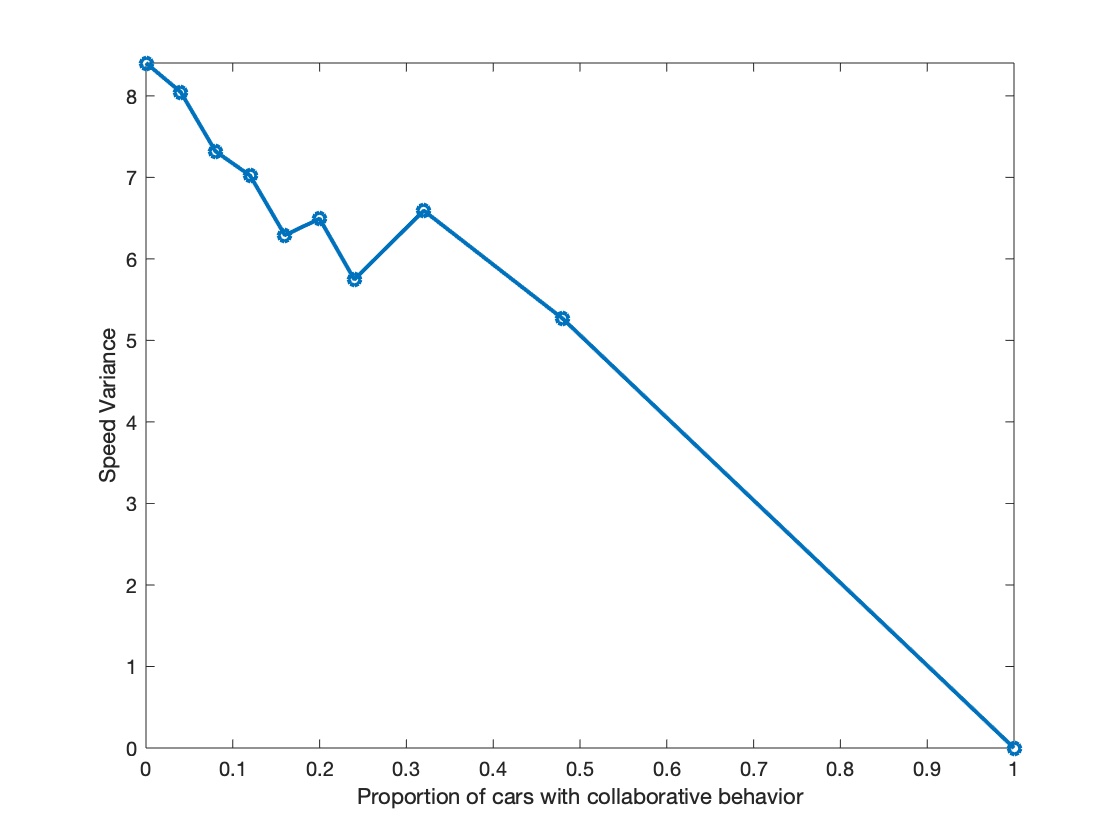}
    \caption{Speed variance with respect to proportion of cars with a collaborative behavior}
    \label{fig:collab}
\end{figure}

\textcolor{black}{In Figure \ref{fig:collab} we present simulations where the proportion of cars $p$ varies from no collaborative behavior to 100\% of drivers following this collaborating behavior. As expected, when $p=0$ (no collaborative behavior) the system has a large speed variance, while when $p=1$ the system is stable and hence the speed variance is close to 0. However, what is interesting to see is that as soon as $p>0$, that is to say as soon as some vehicles starts to have a collaborative behavior, the global speed variance of the system diminishes. In Figure \ref{fig:collab} we ran simulations on a single lane ring-road of $258m$ with $25$ cars during $1000s$ an starting close to the steady-state equilibrium with random initial conditions. Among the $25$ cars the number of cars with a collaborative behavior was 25 ($p=1$), 12 ($p=0.48$), then 8 ($p=0.32$), 6 ($p=0.24$), 5 ($p=0.20$), 4 ($p=0.16$), 3 ($p=0.12$), 2 ($p=0.08$), 1 ($p=0.04$), 0 ($p=0$). For each of these proportions we ran 40 simulations for which we computed the instantaneous spatial speed variance between cars averaged over the last 100s of the simulation, and then averaged it over the 40 simulations. Despite being less effective than control AVs at low penetration rate this observation is an incentive to look more in details at collaborating behaviors. For instance, in these simulations the cars with collaborative behaviors are as evenly distributed in the traffic, and it would be interesting to see if there is any difference when they are clustered.}
\newpage

\section{Future works: macro and mean-field model}
\label{sec:futureworks}

\subsection{Macro-model}
In this paper we consider a micro-model, which gives us a better understanding of the dynamics and behavior of individual cars, and thus a more accurate measure of fuel consumption. However, when the number of cars becomes high, the analysis for optimization and optimal control can become computationally unfeasible. Therefore many macroscopic models (macro-model) have been derived to study the behavior of traffic flow at a larger scale. In these models, the dynamics are distributed and represented by partial differential equations. The first models were scalar, such as the celebrated but limited 
Lighthill-Whitham-Richards 
model, where the density of cars on the road is the only variable and the speed is a decreasing function of density \cite{LighthillWhitham,Richards56}. 
These models regained interest with the emergence of more realistic second-order models \cite{aw2000resurrection,lebacque2007generic,fan2017collapsed}. These models included two equations where both the density and speed were included as variables. The first equation often represents a transport density, while the second equation represents the effect from acceleration. Second order macro-models can also represent traffic waves more easily. One can cite in particular the study of "jamitons" waves \cite{flynn2009self}. A harder question when dealing with macro-models is the question of the interactions between the AVs and the regular traffic flow. While for micro-models, this interaction is relatively easy to represent accurately (one only needs to give the AV a different acceleration law than the other vehicles), the interaction between the AVs and the rest of the traffic flow in a macro-model raises several issues: 
\begin{itemize}
\item Should the AVs also obey there own macro-model and, if so, how are the two macro-models coupled?
\item Should the AVs be represented as individual cars and how should the microscale and macroscale be coupled?
\end{itemize}
One proposition to interact these two models is in the form of an ODE-PDE system, which is given in \cite{delle2014scalar}. Several works were even developed to show that this system makes sense mathematically (i.e. are well-posed) and exhibits the expected behavior \cite{liard2019entropic,liard2019well,HLMP2021}. For the above reasons, we restricted ourselves to microscopic modelling in this paper, even though a macro-scale model may be promising to design efficient controllers that can dissipate traffic instabilities, in particular stop-and-go waves. 

\subsection{Mean-field models}

In this section, we go further and talk about the mean-field models on vehicular traffic. 

Microscopic models describe the details of the traffic flow by studying each individual vehicle's microscopic properties like its position and velocity. The trajectories of the vehicles are predicted by means of ordinary differential equations (ODEs). Macroscopic models, assume a sufficiently large number of vehicles on the road and treats vehicular traffic as fluid flow. In particular, the evolution of the traffic density $\mu$ is governed by partial differential equations (PDEs). Thus, by capturing and predicting the main phenomenology of microscopic dynamics, macroscopic models can provide an overall and statistical view of traffic. One can also use a coupled ODE-PDE system to model the dynamics of a small number of AVs and a large number of regular vehicles on a single lane. This is clearly a combination of the microscopic and macroscopic models using multiple scales together.   

The relationship between the two different scale models, microscopic and macroscopic models, can be both formally and rigorously established via mean-field approach by taking the number of vehicles $N$ to go to infinity.
Let $(x_i, v_i)$ be the position-velocity vector of the $i$-th vehicle and $\mu$ be the density distribution of infinitely many vehicles in the space of position and velocity. The dynamics of the finitely many vehicles can be described by 
\begin{equation}
\label{ODE system}
\begin{cases}
\dot{x}_i = v_i,\\
\dot v_{i} = H*\mu_N(x_i, v_i), \quad i=1, \dots, N,
\end{cases}
\end{equation}
where $H \colon \mathbb{R} \times \mathbb{R}^{+} \mapsto \mathbb{R}$ is a convolutional kernel and $\mu_N(t) = \frac{1}{N}\sum\limits_{i=1}^{N}\delta_{(x_i(t), v_i(t))} $ is a probability measure. The dynamics of the infinitely many vehicles can be described by 
\begin{equation}
\label{PDE}
\partial_t \mu + v \cdot \nabla_x \mu = \nabla_v \cdot [(H*\mu)\mu].
\end{equation}
Furthermore, one can rigorously derive the mean-field limit of the  finite-dimensional ODE system \eqref{ODE system}, and the infinite dimensional mean-field limit \eqref{PDE} (a Vlasov-Poisson type PDE), see \cite{fornasier2014mean}. We also want to point out that the above mean-field approach is different with the so called "mean-field games" approach. For the mean-field games approach, one assumes many agents with perfect knowledge of the system and gives a strategy to solve a game, then passes it to the limit as in the mean-field approach. 

For the multi-lane and multi-class traffic that includes AVs and regular vehicles, one can only consider a mean-field limit for the dynamics of the regular vehicles compared to the finitely many AVs governed by control dynamics. The lane changing maneuvers of the infinitely many regular vehicles lead to a source term of the Vlasov-Possion type PDE. The limit process from a finite-dimensional controlled ODE system to an infinite-dimensional controlled coupled ODE-PDE system can be established in generalized Wasserstein distance. Additionally, one can also consider optimal control problems associated to the controlled ODE and coupled ODE-PDE systems where the cost functions represent, for instance, fuel consumption. Moreover, we have the following theorem, see \cite{fornasier2014mean}. 
\begin{thm}
\label{thm}
The optimal solution to the optimal control problem of the ODE system converges to the optimal solution of the optimal control problem of the coupled ODE-PDE system as the number of regular vehicles $N$ goes to infinity.
\end{thm}
Note that Theorem \ref{thm} implies that one can design controls in the microscopic level and be able to pass the limit to get the control in the mean-field limit level.

\section{Conclusion}
\label{sec:5}
In this paper, we presented a hybrid multi-lane micro-model for traffic flow in a ring-road. This model exhibits stop-and-go waves, and we show that the safety and inventive thresholds in lane changing conditions highly impact the behavior of the system. We use a single AV as a means of control for dissipating traffic instabilities, including stop-and-go waves, and we show that even simple controllers can be very efficient in reducing traffic, whatever the thresholds of the lane changing conditions. Additionally, this can be shown to be a very good basis to derive a controller for mean-field models, when we consider a mean-field limit for the dynamics of infinitely many regular vehicles and control dynamics for finitely many AVs.

\section{Authors contribution statement}
\label{sec:6}
N.K and A.H. performed research and simulations on a project designed by B.P with A.B. and A.H.. S.T, P.A. \textcolor{black}{and T.M.} wrote the simulation code, 
R.D., A.H. \textcolor{black}{and T. M.} helped with the simulations, N.K., A.H., S. M., X. G. and B.P. wrote the paper.

\section{Acknowledgements}
\label{sec:7}
This research is based upon work supported by the U.S. Department of Energy’s Office of Energy Efficiency and Renewable Energy (EERE) under the Vehicle Technologies Office award number CID DE-EE0008872. The views expressed herein do not necessarily represent the views of the U.S. Department of Energy or the United States Government. The authors would also like to thank the C2SMART project. The authors acknowledge the Office of Advanced Research Computing (OARC) at Rutgers, The State University of New Jersey for providing access to the Amarel cluster and associated research computing resources that have contributed to the results reported here. Finally A.H. would like to thank the IEA project SHYSTRA from CNRS.

\bibliographystyle{plain}
\bibliography{Biblio_EPJ}

\end{document}